\documentstyle[prl,aps,psfig,floats]{revtex}
\begin{document}

\title{ \large \bf Statistical properties of fractures in damaged materials.}
\author{Andrea Gabrielli $^{1,2}$, Raffaele Cafiero $^3$ 
and Guido Caldarelli $^4$}

\address{$^{1}$ Dipartimento di Fisica, Universit\`a degli Studi ``Tor Vergata'' 
v.le della Ricerca \\ Scientifica, 1, 00133 Roma, Italy \\
$^2$  I.N.F.M. Sezione di Roma 1, and Dipartimento di Fisica,
Universit\`a di Roma ``La Sapienza'',~Piazzale
A. Moro 2, I-00185 Roma, Italy \\
$^3$ Max-Planck Institut f\"ur Physik komplexer Systeme,
N\"othnitzer Strasse 38, D-01187 Dresden, Germany \\
$^4$ Theory of Condensed Matter Group, Cavendish Laboratory, 
Madingley Road, \\ CB3 0HE Cambridge, UK. 
 }
\maketitle

\begin{abstract}
We introduce a model for the dynamics of
mud cracking in the limit of of extremely thin layers.
In this model the growth of fracture proceeds by selecting 
the part of the material with the smallest (quenched) breaking threshold. 
In addition, weakening affects the area of the sample neighbour to the crack.
Due to the simplicity of the model, it is possible to derive some analytical 
results. In particular, we find 
that the total time to break down the sample grows with the dimension
$L$ of the lattice as $L^2$ even though the percolating cluster 
has a non trivial fractal dimension. Furthermore, we obtain a formula for
 the mean weakening with time of the whole sample. 
\end{abstract}
\vspace{0.5cm}
Among the phenomena leading to formation of cracks in materials \cite{HR}, 
fractures arising from desiccation are widespread both in 
nature  where mud cracks of different length scale are present in the 
landscape \cite{mud1} and in the industry, where paint drying and concrete 
desiccation are intensively studied \cite{mud2,mud3}. 
Despite its relevance, the mechanism and the dynamics of cracking by 
desiccation is, however, still unclear. Recently, an experimental work  
\cite{mudcracks} shed some light onto the problem.  
According to this work, the main source of stress is given by the {\em local} 
friction of the layer of material with the bottom of the container. 
Moreover, the 
characteristic scale of the crack patterns varies linearly with the 
layer thickness.  In the limit of zero thickness {\em the crack patterns loose 
their polygonal structure} (the characteristic size of the polygons is zero) 
and become {\em branched fractals}.

In this letter we introduce an extremely simplified lattice model for 
cracking of thin mud (or paint) layers, inspired by the 
vectorial and scalar models described in \cite{CDP,ZVS} 
and by invasion percolation\cite{IP}, for which a theoretical approach, 
based on the Run Time Statistics (RTS) scheme \cite{RTS}, allows us to compute 
some relevant quantities, such as the evolution of the breaking probability. 
Many cellular automata models for crack propagation present 
as a basic ingredient a non-local Laplacian field 
(electric field, electric current) which 
drives the formation of the cracks, acting on a solid network of bonds or 
sites \cite{ZVS}. 
In others the stress field evolves by keeping minimum the energy of the system.
In such a case the components of the vectorial equations obtained are 
similar to the equations describing the action of a Laplacian field \cite{CDP}.

In this model no explicit Laplacian 
field is present, since the stress is represented by a {\em local} random 
breaking threshold. At each time step, the bond with the smallest 
threshold is removed from the lattice. Short ranged correlations are 
introduced through a damaging of the thresholds of the bonds 
nearest neighbours of the bond removed. In defining this model we are 
driven by the above cited experimental observation \cite{mudcracks} that in 
an extremely thin layer of mud or paint the only source of stress 
is the {\em local} friction with the container. No globally applied  
Laplacian field seems to be present. Moreover, since the drying mud is a 
mixture of a liquid and a solid (usually amorphous) phase, no long range 
stress relaxation is present, although the growing crack can 
affect the properties of the medium in its vicinity. The simplifications 
with respect to Refs. \cite{CDP,ZVS} restrict the 
applicability of the model to the special cases above mentioned, yet it allows 
one to write down explicitly some equations for the evolution of the breaking 
probability.

The model is defined as follows: On a square 
lattice a quenched random variable $x_i$ is assigned to 
each bond $i$ where the $x_i$'s are extracted from
a uniform distribution between $0$ and $1$. The bond with the 
lowest value of the variable is selected and removed. Then damage is applied, 
and the bonds nearest neighbours are weakened, i.e. a new threshold $x'_i$ is 
extracted with a uniform probability between $0$ and the former value
$x_i$. This should mimic the stress enhancing nearby crack tips \cite{HR}. 
Then, the next bond with the minimum value of threshold is removed from the 
system and a new weakening occurs. The process continues until a percolating 
cluster of fractures divides the sample in two separate parts. 
>From the point of view of mud cracking, the 
two-dimensional lattice represents a very thin layer of mud (paint), and 
quenched disorder {\em accounts for local stress induced by inhomogeneous 
desiccation of the sample}. 
The dynamics is assumed to be quasi-static,  since we assume
evolution of cracks in mud desiccation to be a slow process.
Some authors  correctly point out that otherwise 
time-dependent effects and a non-equilibrium dynamics are relevant 
in crack propagation \cite{sornette}.

In this model we have eliminated the explicit presence of an external field
(applied stress) and of the response of the material
(strain of bonds). The only quantity present is the breaking threshold, whose
dynamics is chosen to reproduce the evolution of cracks. This simulates 
the presence of a local stress field, acting not on the boundaries but directly
on each bond. Our assumption is based on the experimental results in ref. 
\cite{mudcracks}, where as the mud layer becomes thinner only the 
inhomogeneities drive the nucleation of cracks. Furthermore the hypothesis of
crack developing under the same state of strain not only is usually applied
in the presence of thermal gradients \cite{mud4}, but is also commonly reported
in experiments of loading of softened material \cite{mud2}.
Hence, such a model is 
particularly suitable to describe, for example, paint drying, where the stress
applied to the painted surface depends on the local action of external
condition (density gradient in the paint).

\begin{figure}[h]
\begin{minipage}[l]{6.0cm}
\vspace{-0.4cm}
\psfig{file=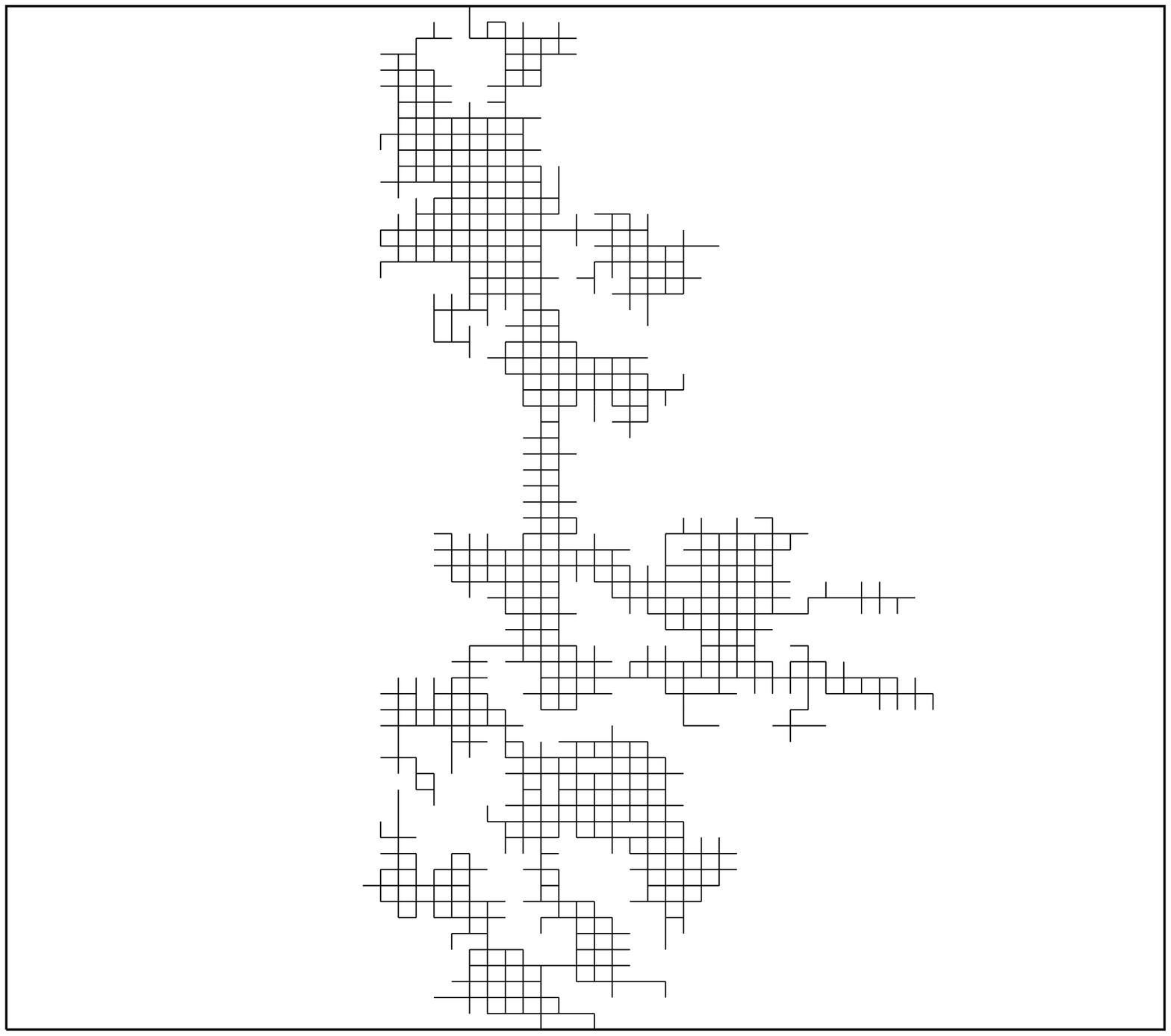,height=5.5cm,angle=-90}
\vspace{0.25cm}
\caption{A typical spanning fracture obtained with the model.  
In the simulations we deal with periodic boundary condition on the sides, 
i.e. left and right boundary coincide.}
\label{fig1}
\end{minipage}
\begin{minipage}[r]{7.5cm}
\psfig{file=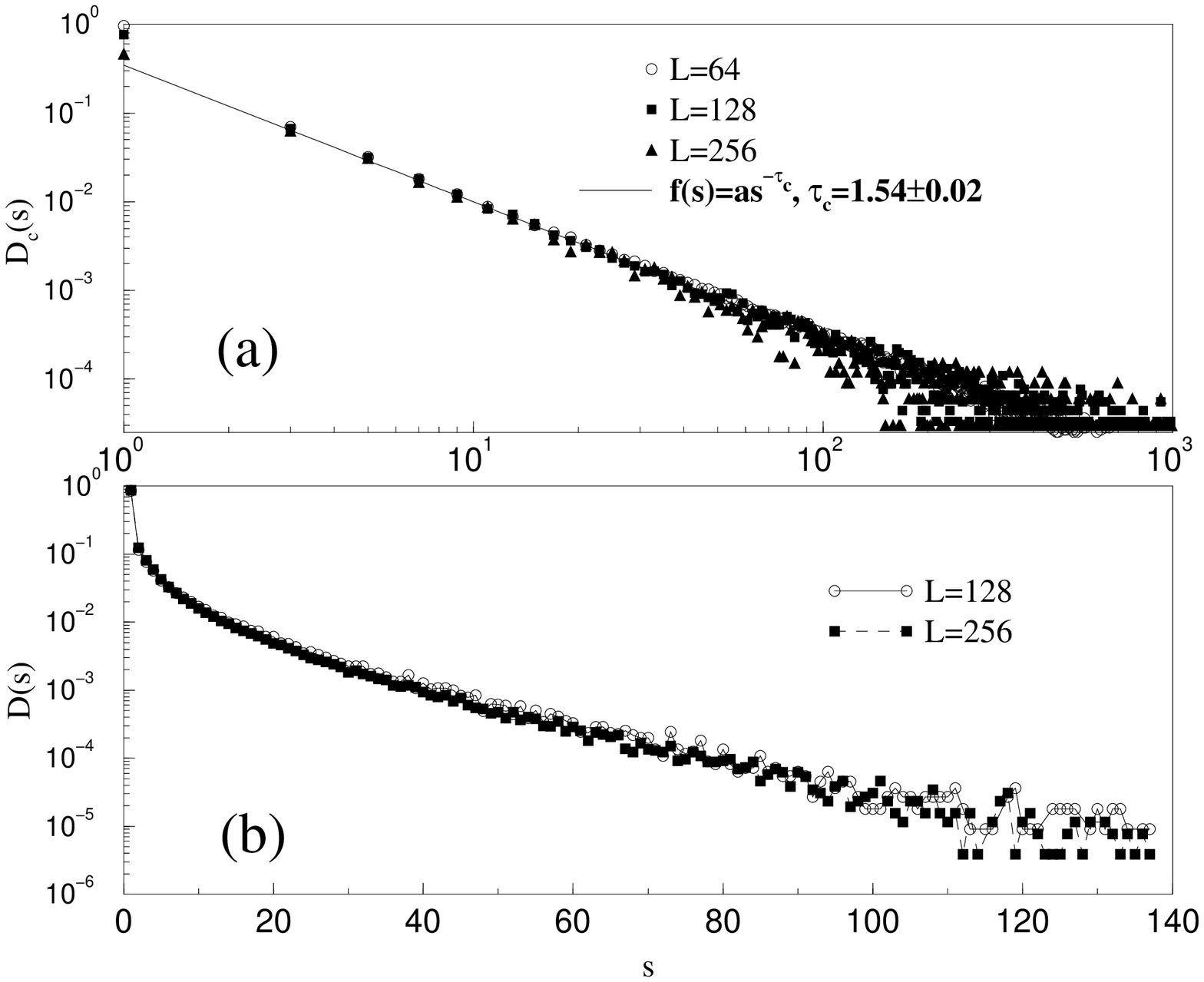,height=6.0cm,angle=0}
\vspace{-0.4cm}
\protect\caption{(a) Probability distribution ($log_{10}$-$log_{10}$ plot) $D_c(s)$ 
of the size of the clusters vs the size, for $L=64,128,256$ 
together with a least square 
fit. (b) Avalanche size distribution (linear-$log_{10}$ plot) $D(s)$ 
for $L=128,256$. While the cluster distribution is power law, 
the avalanche distribution is exponential.}
\label{fig2}
\end{minipage}
\end{figure}

Despite the simplicity of the model rule, the findings are rather interesting. 
We have performed numerical simulations, with cylindrical symmetry 
(periodic boundary conditions in the horizontal direction)  
for various system sizes $L$. The dynamics stops 
as soon as a crack spanning the system in the vertical direction appears.
In Fig.\ref{fig1} a typical percolating cluster is shown.
The fractal dimension of the percolating cluster is 
computed with a box-counting method. We find $D_f=1.77 \pm 0.02$ for 
all values of $L$. We restrict the analysis to the spanning cluster to 
reduce the finite size effects present for the smaller clusters.
Also the distribution of finite clusters is not trivial,  
showing a clear power law with exponent $\tau_c=1.54\pm0.02$ 
(see Fig.\ref{fig2}a).

A recent interest has focused on acoustic emission in order to 
understand if the power law observed both in 
experiments\cite{PPVAC} and in models\cite{CDP,ZVS} is related to
 Self-Organized Criticality
(SOC)\cite{BTW}. The presence of SOC would mean that the dynamics 
of fractures leads the sample to a steady state where small 
variation of the external field can trigger reaction at any length-scale.
In particular the external field in this case is the applied stress, and the
response of the sample can be considered as the energy released (acoustic 
emission) 
by one avalanche of cracks, where avalanche means a causally connected 
series of breakdowns.
In this over-simplified model the external stress can be considered constant, 
since the only change after any single breakdown is the damaging of the 
nearest neighbours.
Then we monitored, as a measure for the acoustic emission 
the size of an avalanche which is defined as follows.  
Let us suppose that a bond $i$ grows (i.e. it is broken) 
at time $t$; this is the {\em initiator} 
of an avalanche, which is defined as the set of events geometrically 
and causally connected to the initial one (bond $i$). ``Causal" 
connection refers to the weakening following any bond breaking.
In particular, when bond $i$ grows at time $t$, the avalanche goes on at time 
$t+1$ if a non-broken first neighbour bond $j$
of $i$ is removed. 
At time $t+2$ the avalanche goes on if a bond $k$ grows where $k$ is a 
non-broken first neighbour of $i$ or $j$ and so on.
In FIG.\ref{fig2}b we show a linear-log plot of the probability 
distribution of avalanche size, versus sample size $L$.  
We get, after a power law transient, an exponential distribution 
signalling that a characteristic size exists for the avalanches. 
This result for avalanches is similar to those obtained for a scalar model of 
Dielectric Breakdown, but differs from the avalanche 
behaviour in models of fracture \cite{CDP,ZVS}.
The reason for that is twofold. Firstly, in the present 
definition of an avalanche the threshold
is changed only for the nearest neighbours. This introduces a 
typical length scale, while
other definitions consider as the threshold the ratio between 
local field and resistivity, thus
giving the possibility of large scale correlations. 
Secondly, in this model grown bonds {\em are removed from the system}. 
This represents a substantial difference with the SOC models presented in the 
literature.
For example, in a simple toy model of SOC due to Bak and Sneppen\cite{BS}
(where a similar refresh for the threshold is present) 
the dynamics produces clear power laws in the 
avalanche distribution. 
There, each site (species) deleted is replaced by a new one
and is not definitively removed.
This is a crucial point, since indeed power law behaviour in the presence of a 
scalar field seems to be related to a ``reconstructing rule'' that allows one to 
deal with a system where bonds are not removed but they change their status. Then only in the case of plastic deformation, one is in the presence of a steady state, as correctly pointed out by Ref.\cite{ZVS}.

Since the evolution of the crack is taken to be extremal, i.e. the bond with
the minimal threshold is removed from the system, one is allowed to apply a 
recent theoretical tool, the {\em Run Time Statistics} (RTS) \cite{RTS},
to this model in order to predict 
the probability of breakdown of one bond and then the
resistance of the whole medium.
RTS has been introduced to map deterministic 
quenched extremal dynamics to a stochastic description. 
As discussed in \cite{RTS} we can associate with every bond $j$ an 
effective probability density $p_{j,t}(x)$ for the threshold $x_j$ 
conditioned by the growth history.
We can write the growth probability $\mu_{i,t}$ of the bond $i$ at time $t$ as
\begin{equation}
\mu_{i,t}=\int_0^1 dx\,p_{i,t}(x) \left[ \prod_{k(\neq i)}^{\partial C_t} 
\int_x^1 dy \, p_{k,t}(y) \right] .
\label{mu}
\end{equation}
In this case (contrary to the case of Invasion Percolation), 
$\partial C_t$ is the whole set of non-broken bonds. 
Note that at time $t$ the number of bonds in $\partial C_t$ is 
$(2L^2 - t)$, i.e., the total number $2L^2$ of bonds in a square lattice 
with side $L$ minus the number of broken bonds before time $t$. 
Now we can update every $p_{j,t}(x)$ by conditioning them to 
the latest growth event, this giving the $p_{j,t+1}(x)$'s. 
We call $m_{i,t+1}(x)$ the effective probability density at time $t+1$ of the 
latest grown bond $i$. It is given by
\begin{equation}
m_{i,t+1}(x)=\frac{1}{\mu_{i,t}} p_{i,t}(x)\left[ \prod_{k(\neq i)}
^{\partial C_t} \int_x^1 dy\,p_{k,t}(y) \right] .
\label{m}
\end{equation}

For the remaining bonds we have to distinguish between the weakened ones 
(the nearest neighbours of $i$) and the rest.
For the latter we have
\begin{equation}
p_{j,t+1}(x)=\frac{1}{\mu_{i,t}} p_{j,t}(x)\!\!\int_{0}^{x}\!\! dy \,p_{i,t}(y) 
\!\!\left[ \prod_{k(\neq i,j)}^{\partial C_t}\!\!
\int_y^1 \!\!dz \, p_{k,t}(z) \right] 
\label{p1}
\end{equation}

while for the former we obtain
\begin{eqnarray}
p_{j,t+1}(x)&=&\int_{0}^{1}dy\, \frac{1}{y} \theta(y-x)\left\{\frac{1}
{\mu_{i,t}} p_{j,t}(y)\int_{0}^{y} dz \, p_{i,t}(z)\times\right.\nonumber \\
&\times&\left.\left[ \prod_{k(\neq i,j)}^{\partial C_t}
\int_x^1 du \, p_{k,t}(u) \right]\right\}
\label{p2}
\end{eqnarray}

Eqs. (\ref{mu}, \ref{m}, \ref{p1}) coincide with the 
ones introduced for the RTS approach
to Invasion Percolation (apart from the different definition of the 
growth interface $\partial C_t$). Eq. (\ref{p2}), instead, is new and refers
to the nearest neighbours weakening.
Eqs. (\ref{mu}, \ref{m}, \ref{p1}, \ref{p2}) allow one to study the extremal 
deterministic dynamics as a stochastic
process. In particular, $\mu_{i,t}$ can be used to evaluate 
systematically the statistical weight of a fixed growth path, while 
$p_{j,t}(x)$ stores information about the growth history.
A very important quantity to characterize the properties of the dynamics
is the empirical distribution (or histogram) of non-broken thresholds.
This quantity is defined as:
\begin{equation}
h_t(x)=\sum_{j \in \partial C_t} p_{j,t}(x)
\end{equation}
where, $h_t(x)dx$ is the number of non-broken bonds between $x$ and  $x+dx$
at time $t$. 

Considering the effect of the growth of bond $i$ at time $t$ on this
quantity,
\begin{equation}
\!\!h_{t\!+\!1}(x)\!=\!h_t(x)\!-\!m_{i,t\!+\!1}(x)\!-\!\!\sum_{j(i)}
p_{j,t}(\!x)\! +\!\sum_{j(i)}p_{j,t+1}\!(\!x)
\label{zappinbianc}
\end{equation}
where $j(i)$ indicates the set of non-broken bonds $j$ nearest neighbours of 
$i$ and $m_{i,t+1}(x)$ and $p_{j,t+1}(x)$ are given respectively by 
Eq.(\ref{m}) and Eq.(\ref{p2}).
In order to evaluate the statistical properties of the crack evolution, 
we averaged over all the possible path growths until time $t+1$.
Introducing the notation $\left\langle...\right\rangle$ for this average, the l.h.s. of 
Eq.(\ref{zappinbianc}) can be computed as
\begin{equation}
\left\langle h_{t+1}(x)\right\rangle =\!||\partial C_{t+1}||\phi_{t+1}(x)\!=
\![N\!-\!(\!t\!+\!1\!)]\phi_{t+1}(x), 
\label{zapp}
\end{equation}
where $N=2L^2$ is the total number of bonds in the lattice and 
$\phi_t(x)$ represents the average thresholds distribution over
the non-broken bonds at time $t$ (normalized to $1$).
For the r.h.s. of Eq.(\ref{zappinbianc}) the main difficulty arises in 
the evaluation of $\left\langle m_{i,t+1}\right\rangle$ and 
$\left\langle\sum_{j(i)} p_{j,t+1}(x)\right\rangle$.
Following \cite{RTS} we can write 
\begin{equation}
\left\langle m_{i,t+1}\right\rangle\simeq (N-t)\phi_t(x)e^{-(N-t)\int_0^x
\!dy\,\phi_t(y)}.
\label{mind}
\end{equation}
Assuming $x$ to be independent of the number $n_t$ of bonds weakened 
at time $t$ and applying the same mean field approximations used 
to obtain Eq.(\ref{mind}) \cite{RTS} we have
\begin{equation}
\left\langle\sum_{j(i)} p_{j,t}(x)\right\rangle=n_t\phi_t(x),
\end{equation}
while for the annealed bonds, we 
obtain, after some algebra, the expression
\begin{equation}
\left\langle\sum_{j(i)} p_{j,t+1}(x)\right\rangle= 
\frac{n_t(N\!-\!t)}{N\!-\!t\!-\!1}\!\int_x^1 \!\!\!dy\, \frac{\phi_t(y)}{y}
[1\!\!-\!\!e^{-(\!N\!-\!t\!)\int_0^y dz\phi_t(z)}].
\end{equation}
Then we can write the following equation for the $\phi_{t+1}(x)$:
\begin{eqnarray}
\phi_{t+1}(x)&\!\!=\!\!&\frac{(\!N\!-\!t\!-\!n_t)}{N\!-\!t\!-\!1}
\phi_t(x)\!-\!\frac{\!N\!-\!t\!}{N\!-\!t\!-\!1}\phi_t(x) 
e^{-(N-t)\!\int_0^x dy\phi_t(y)}+ \nonumber \\
&\!+\!&n_t\frac{\!N\!-\!t\!}{(\!N\!-\!t\!-\!1\!)^2}\int_x^1dy\!
\frac{\phi_t(y)}{y} [1-e^{-(N-t)\int_0^y dz \phi_t(z)}]
\label{equazione}
\end{eqnarray}
Note that even at percolation time $N-t$ is a large number. For this reason 
terms  in eq.(\ref{equazione}) which contain the exponentials are 
negligible for those $x$ for which $\int_0^x dy\phi_t(y)$ is finite 
(larger than $1/(N-t)$).
It is easy to show that the continuum limit of Eq.(\ref{equazione}) for 
such values of $x$
is invariant under the rescaling $L\rightarrow aL$ (i.e. $N \rightarrow 
a^2N)$ and $t\rightarrow a^2t$. 
This result is based on the assumption that:
\begin{equation}
n_t(L)=n_{a^2t}(aL)
\label{n}
\end{equation}
 The numerical simulations suggest the following 
scaling form for $n_t(L)$:
\begin{equation}
n_t(L)=n_{max}\left[\frac 1{1+t/AL^2}\right]^\beta ,
\label{nt}
\end{equation}
where $\beta=0.23\pm0.02$ and $n_{max}=6$ is the lattice coordination number.
This form for $n_t$ satisfies Eq.(\ref{n}).  

\begin{figure}[h]
\begin{minipage}[l]{6.8cm}
\vspace{-0.25cm}
\psfig{file=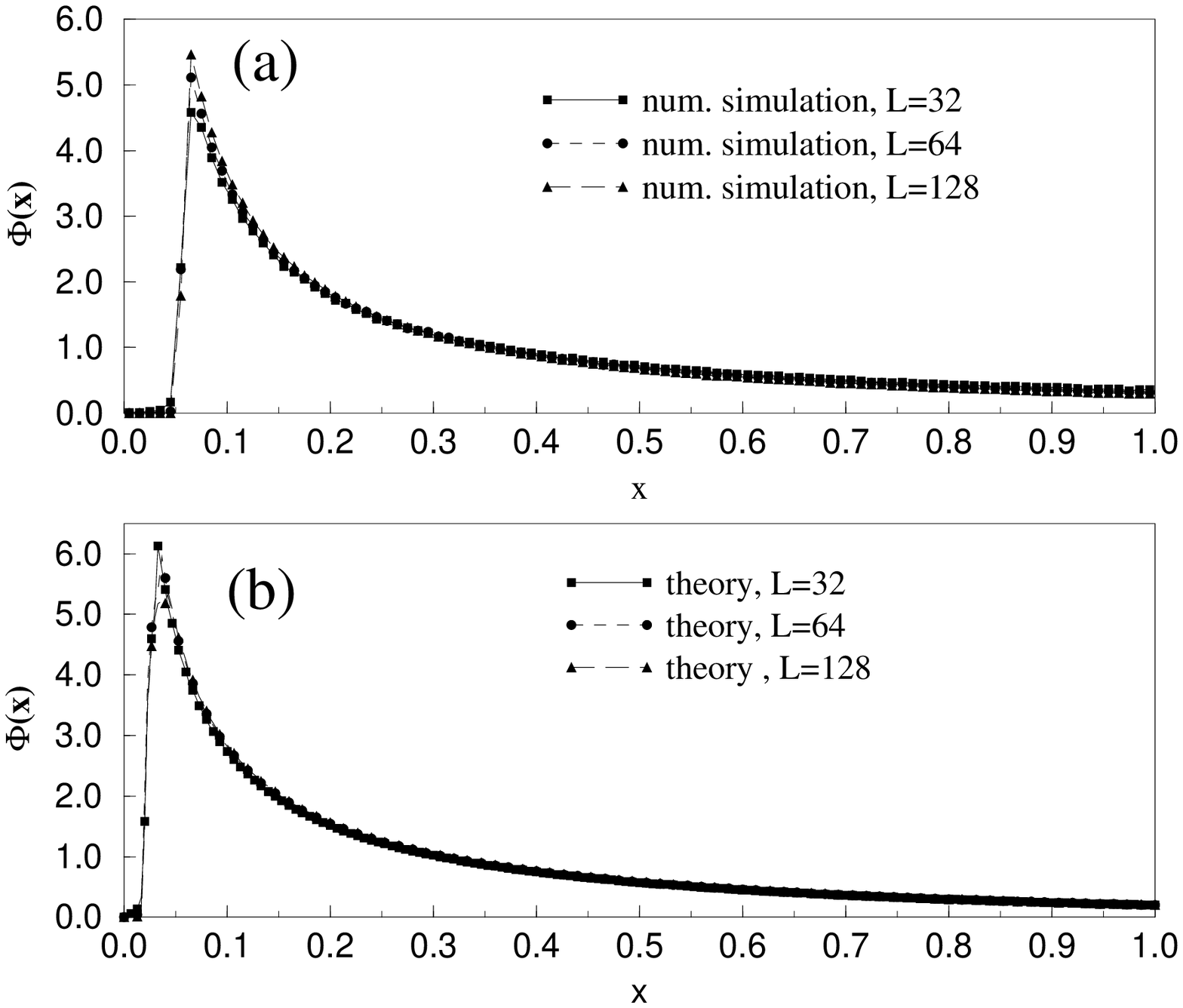,height=6.0cm,width=6.8cm,angle=0}
\caption{Solution of Eq.\ref{equazione} for the histogram 
$\phi_t(x)$ at the spanning time (b), compared with simulations (a).} 
\label{fig3}
\end{minipage}
\hspace{0.2cm}
\begin{minipage}[r]{6.8cm}
\vspace{0.2cm}
\psfig{file=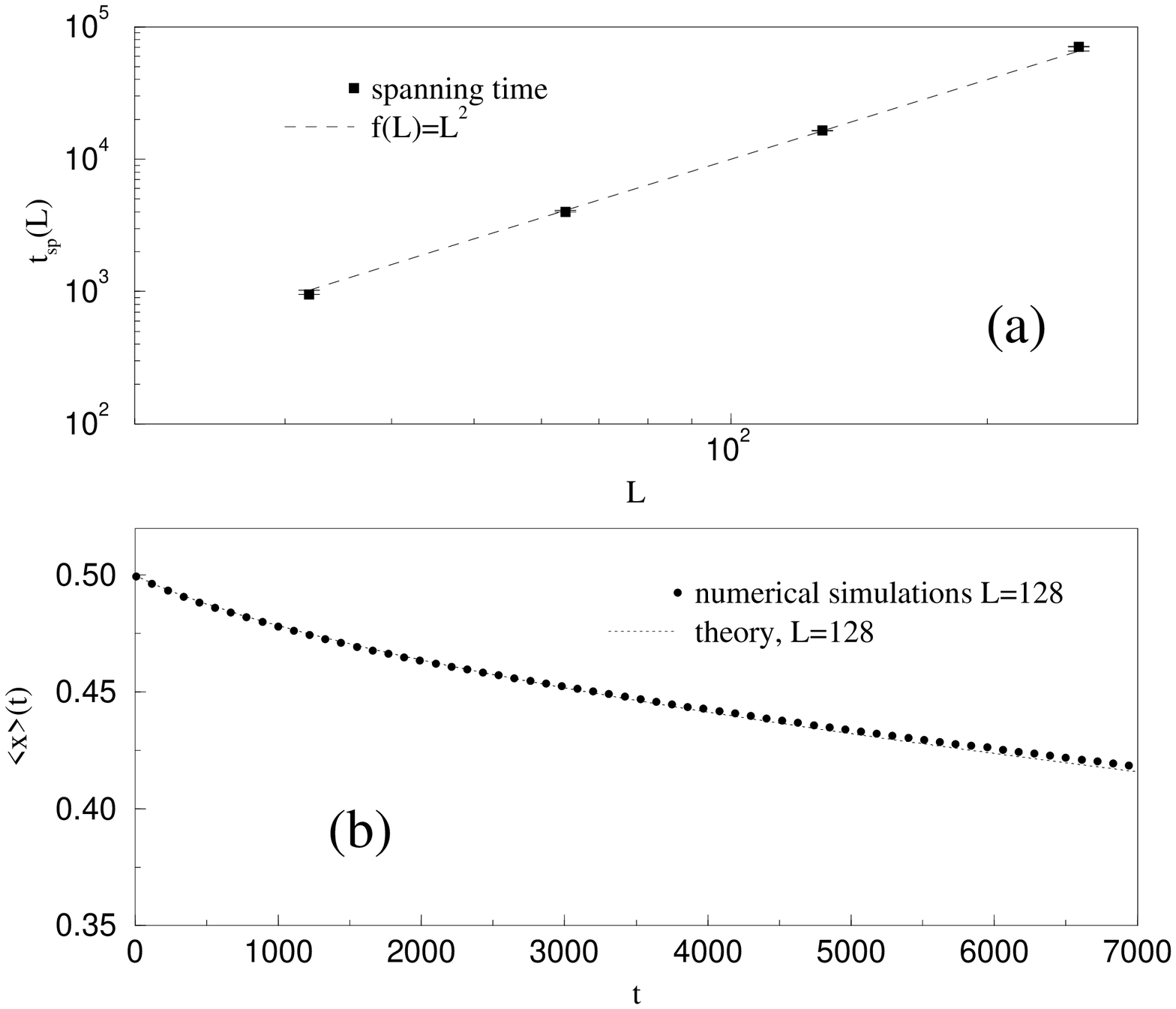,height=6.0cm,width=6.8cm,angle=0}
\caption{(a) Spanning time versus system size $L$. One can see
 a nice agreement with the expected scaling law $t_{sp}(L)\propto L^2$.
(b) Solution of eq.\ref{xmed} compared with numerical simulations.}
\label{fig4}
\end{minipage}
\end{figure}

We are then able to make three important predictions:
Firstly, we find both analytically, from the numerical solution of 
Eq.\ref{equazione}, 
and from computer simulations a 
discontinuity in the histogram (see Fig.\ref{fig3}), signalling that the i
system evolves in a way such as to remove all bonds with threshold 
smaller than some critical value.
Secondly, from the symmetry properties of Eq. \ref{equazione} we deduce 
that the number $t_{sp}(L)$ of broken bonds at the percolation time 
is proportional to $L^2$, even though the percolating cluster is fractal. 
This result is deduced supposing that at the percolation time the shape of 
the histogram is independent of $L$ which is verified by numerical 
simulations (Fig.\ref{fig3}-a).
The numerical check of this result is 
presented in Fig.\ref{fig4}a.
Thirdly, we present a qualitative result for the average value of the 
thresholds (i.e. the resistance of the material) $\left\langle 
x\right\rangle (t)$.
For discretized times we find:
\begin{equation}
\langle{x}\rangle(t+1)=\left(1-\frac{n_t-2}{2(N-t-1)}\right)
\langle{x}\rangle(t)+\frac{1+n_t/[2(N-t)]}
{N-t-1} 
\times \int_0^1dx e^{-(N-t)\int_0^x dy \phi_t(y)},
\label{xmed}
\end{equation}
 
It is possible to show that $\langle{x}\rangle(t+1) < \langle{x}\rangle(t)$ 
unless $\langle{x}\rangle(t)$ is exponentially small in $N-t$. 
T0his means that on average the medium weakens during the evolution
even if the weakest bond is removed at any time step.
In Fig.\ref{fig4}b we show the time evolution of
 $\langle{x}\rangle(t)$ obtained from 
computer simulations, compared with 
the solution of eq. (\ref{xmed}). Our analytical 
results are in nice agreement with numerical simulations.

As a conclusion, we would like to point out that, to our knowledge, 
apart the qualitative results of \cite{mudcracks}, 
no quantitative experimental results are available. 
For example, a measurement of the fractal dimension of 
cracks or their size distribution would be extremely useful to further test 
the predictions of this model.
At the moment, this model seems able to capture, with its extremely simplified 
dynamics, some basic properties of fracturing processes.

In conclusion we have presented a new model for fractures, that 
we believe should be applicable to the case of drying paint and mud cracking, 
for extremely thin samples. Due to its extreme simplicity,  
the model is particularly suitable for large scale simulations and takes into
account the damaging effects involved in fracture propagation. 
Even in this simple model we are able to analyze 
which conditions trigger SOC behaviour in such systems. 
Furthermore, the change in the threshold distribution induced by the 
annealing allows us to write down explicitly the form of the breakdown 
probability for the bonds of the sample. In the future we plan to extend 
the analytical study of the model to the computation of its the relevant 
critical exponents. This is in principle possible by combining the RST 
method with the fixed scale transformation approach (FST) \cite{fst}.
The authors acknowledge the support of the EU grant Contract No. FMRXCT980183.
GC acknowledges the support of EPSRC.


\vskip-20pt
\begin{thebibliography}{99}

\bibitem{HR} H. J. Herrmann, S. Roux 
{\em Statistical Models for the Fractures of Disordered Media, 
North Holland} 1990.

\bibitem{mud1} J. T. Neal, {\em Geol. Soc. Am. Bull.} {\bf 76} 1075, 1966; 
J. T. Neal and W. S. Motts, {\em J. Geol.} {\bf 75} 511, 1967;  
G. Korvin, Pure. Appl. Geophys. {\bf 131} 289, 1989.

\bibitem{mud2} J. Walker, {\em Sci. Am.} {\bf 255} 178, 1986. 

\bibitem{mud3} Wittmann F. H., Slowik A., Alvaredo A. M., 
{\em Materials and Structures} {\bf 27} 499, 1994;
Nishikawa T., Takatsu M. {\em Cement and Concrete Research} {\bf 25} 1218,  
1995.

\bibitem{mudcracks} A Groisman and E. Kaplan, {\em Europhys. Lett.} 
{\bf 25} 415, {1994}.

\bibitem{CDP} G. Caldarelli, C. Castellano, A. Vespignani {\em Phys. Rev. E} 
{\bf 49} 2673, 1994; G. Caldarelli, F. D. Di Tolla, A. Petri
{\em Phys. Rev. Lett.} {\bf 77} 2503 1996.

\bibitem{ZVS} S. Zapperi, A. Vespignani, H. E. Stanley
{\em Nature} {\bf 388} 658, 1997.

\bibitem{IP} D. Wilkinson, J. F. Willemsen
{\em J. Phys A (London)} {\bf 16} 3365, 1983.

\bibitem{RTS} {M. Marsili}
{\em J. Stat. Phys.} {\bf 77} 773, 1994;
A. Gabrielli, R. Cafiero, M. Marsili and L. Pietronero
{\em Europhys. Lett.} {\bf 38} 491, 1997.

\bibitem{sornette} D. Sornette, C. Vanneste
{\em Phys. Rev. Lett.} {\bf 68} 612, 1991.

\bibitem{mud4} {Bazant Z. P., Carol I. (eds)}, {\em Creep and Shrinkage of
Concrete} E \& FN Spon 1993. 

\bibitem{PPVAC} {A. Petri, G. Paparo, A. Vespignani, A. Alippi, 
M. Costantini} {\em Phys. Rev. Lett.} {\bf 73} 3423, 1994.

\bibitem{BTW} {P. Bak, C. Tang, K. Wiesenfeld}
{\em Phys. Rev. Lett.} {\bf 59} 381, 1987.

\bibitem{BS} {P. Bak, K. Sneppen}
{\em Phys. Rev. Lett.} {\bf 71} 4083, 1993.

\bibitem{fst} {A. Erzan, L. Pietronero and A. Vespignani} 
{\em Rev. Mod. Phys. } {\bf 67} 545, 1995.


\end{thebibliography}
\end{document}